\documentclass{bmcart}

\usepackage{graphicx}
\usepackage{caption}
\usepackage{subfigure}
\usepackage{amsthm, amsmath, amssymb}
\usepackage{mathrsfs}
\usepackage{url}
\usepackage{float}

\begin{document}
\begin{frontmatter}
\begin{fmbox}
\dochead{Research}
\title{The L\'{e}vy Flight of Cities: Analyzing Social-Economical Trajectories with Auto-Embedding}
\author[
   addressref={aff1},                   
]{\inits{LFT}\fnm{Linfang} \snm{Tian}}
\author[
   addressref={aff2},
]{\inits{KZ}\fnm{Kai} \snm{Zhao}}
\author[
    addressref={aff1},
]{\inits{JMY}\fnm{Jiaming} \snm{Yin}}
\author[
    addressref={aff3},
]{\inits{HV}\fnm{Huy} \snm{Vo}}
\author[
   addressref={aff1},
   corref={aff1},
   email={wxrao@tongji.edu.cn}
]{\inits{WXR}\fnm{Weixiong} \snm{Rao}}
\address[id=aff1]{                            
  \orgname{School of Software Engineering, Tongji University}, 
  \street{Caoan Road},                        %
  \postcode{201804}                           
  \city{Shanghai, China},                            
  \cny{\\ \ddag Linfang Tian and Kai Zhao contributed equally to the paper}                                 
}
\address[id=aff2]{
  \orgname{Robinson College of Business, Georgia State University},
  \street{Gilmer Street},
  \city{Atlanta},
  \cny{USA}
}
\address[id=aff3]{
    \orgname{the City College of the City University of New York, and the Center for Urban Science and Progress, New York University},
    \city{New York},
    \cny{USA}
}
\end{fmbox}

\begin{abstractbox}
\begin{abstract} 
It has been found that human mobility exhibits random patterns following the L\'{e}vy flight, where human movement contains many short flights and some long flights, and these flights follow a power-law distribution. In this paper, we study the social-economical development trajectories of urban cities. We observe that social-economical movement of cities also exhibit the L\'{e}vy flight characteristics. We collect the social and economical data such as the population, the number of students, GDP and personal income, etc. from several cities. Then we map these urban data into the social and economical factors through a deep-learning embedding method Auto-Encoder. We find that the social-economical factors of these cities can be fitted approximately as a movement pattern of a power-law distribution. We use the Stochastic Multiplicative Processes (SMP) to explain such movement, where in the presence of a boundary constraint, the SMP leads to a power law distribution. It means that the social-economical trajectories of cities also follow a L\'{e}vy flight pattern, where some years have large changes in terms of social-economical development, and many years have little changes. 

\end{abstract}


\end{abstractbox}
\end{frontmatter}

\section*{Introduction}

Urban studies seek to understand and explain regularities observed in the world's major urban systems. Cities are complex systems \cite{chroneer2019urban, slach2019urban, frick2018big, jiao2020assessment, alves2018crime, ozdemir2020distributional,topirceanu2018weighted, beare2020emergence} with many people living in and complex relationships among various factors. Previous works have studied the mobility of people \cite{corral2020truncated} and show that the movement of human society is statistically random. 
A lot of studies are about the rank-order of cities \cite{gonzalez2020spanish, luckstead2017pareto, luckstead2017size,wu2019transit,frick2018big,bee2019distribution}, Pareto law \cite{luckstead2017pareto,toda2017note} and Zipf's law \cite{wei2021characteristics,jiao2020assessment,sun2021did}. In this paper, we study the development trajectories of cities to contribute to the sustainability and innovation of cities \cite{slach2019urban,bibri2019big,visvizi2018policy, duran2017sustainability, bibri2019big}. This paper will aid policymakers, city planners and government officials to understand the nature of urban development and design a sustainable smart cities using computational social science models.

In this paper, we follow the urban dynamics model above and assume that cities move in two directions \cite{bharath2018modelling}: one is economic growth, the other is the development of social civilization. We study the datasets of four Asian cities: two in China including Hong Kong and Shanghai, the third is Singapore, the fourth is Tokyo, Japan. (see Table \ref{tab:4citiesdatasets}) All of them have economic factors such as GDP, GDP of secondary industry and GDP of tertiary industry, and social factors such as population, education and publication. It covers the most commonly used data types for measuring urban development \cite{gonzalez2020spanish}. Firstly, it is clear that the research object is urban mobility \cite{jiang2018deepurbanmomentum}, namely {the change amount of urban economic and social development}, and the {change amount of social and economic factors} is obtained. Then, we apply the recently popular artificial neural network embedding technique, namely Auto-Encoder \cite{bengio2007greedy}, on all economic factors to extract a low-dimensional latent vector. Min-max normalization is performed on the data first, and the same was done on the data of social factors. Next, we determine the step size distribution of the city movement. According to Akaike Information Criterion \cite{sakamoto1986akaike}, the distribution model is fitted to get the optimal probability distribution. The results show that the movement of Hong Kong is more in line with the truncated power law \cite{chen2019power} distribution, Shanghai city and Tokyo move more power law, and Singapore moves in the pattern of exponential distribution. To the best of our knowledge, this article is the first work that examines the movement of urban social-economical developments using computational social science models and explain the generalization model behind it.

The contribution of this paper is as follows. First, we extract the increment distribution function of city's society according to city's economy. 
Second, we demonstrate that log-normal processes \cite{feng2020accumulative,montebruno2019tale,newberry2019self} in the presence of a boundary constraint, approximately yields a generative process with a power law distribution. This result is a step towards explaining the emergence of L\'{e}vy flight patterns in city development. Thirdly, we use the stochastic multiplicative processes \cite{sornette1997convergent, mitzenmacher2004brief,guerrero2020multiplicative,hodgkinson2020multiplicative,zanette2020fat,fenner2018multiplicative} to explain the urban development trajectory, regarding city as an organism growing theory \cite{shultz2021natural}.

\section*{Results} 
\textbf{Power-law fit for city trajectory flight.} 
First, we draw the social-economical trajectories (see Figure \ref{fig:Anto-Encoder}) of cities and get the histograms (see Figure \ref{fig:histogram}) of walk lengths. We fit the walk length distribution (see Figure \ref{fig:Power-law}) of the Shanghai city, the Hong Kong City, Singapore and the Tokyo city. We fit truncated power law \cite{corral2020truncated}, log-normal, power law \cite{mitzenmacher2004brief,feng2020accumulative,sakiyama2021power,chen2019power,pang2019nonstationary}, and exponential distribution \cite{miyaguchi2019brownian}. (see Table \ref{tab:pdf}) Then use Akaike weights (see Table \ref{tab:Akaikeweights}) to choose the best fitted distribution. We find that the urban development step size of Hong Kong fits Truncated power-law distribution with $\alpha = 1.3547$, and the walk distributions of the other three cities fit Power-law distributions. The exponent $\alpha$ is  2.2829 for Shanghai, 2.6075 for Singapore and 2.6016 for Tokyo. Assuming that urban development satisfy the stochastic multiplicative processes, we draw the walk length change rate (see Figure \ref{fig:change rate}) and logarithm of change rate (see Figure \ref{fig:logchangerate}). The deducted exponents $\alpha$ by SMP are similar to those of the fitted values of $\alpha$. (see Table \ref{tab:estimatedparameters})

\textbf{Mechanisms behind the Power law pattern.} 
A city should be considered an ever changing organism instead of a static one. At each step $t$, the organism may grow or shrink \cite{slach2019urban}, according to a random variable $R_{t}$, so that the change of the city $l_{t}=r_{t-1}l_{t-1}$. This is stochastic multiplicative processes \cite{guerrero2020multiplicative} $l_{t}=r_{t}r_{t-1}...r_{1}l_{0}$. The idea is that the random growth of an organism is expressed as a percentage of its current increment, and is independent of its current actual size. Then we find
\begin{eqnarray}
    \ln l_{t} &=& \sum_{i=1}^t \ln r_{i}+ \ln l_{0}
\end{eqnarray}

Assuming the random variables $\ln R_{i}$ satisfy independent and identical distributions with mean $v$ and variance $D$, the Central Limit Theorem says that $\ln L_{t}=\sum_{i=1}^t \ln R_{i} + \ln l_{0}$ converges to a normal distribution with mean $vt$ and variance $ Dt $ for sufficiently large $t$, which means $L_{t}$ converges to a log-normal distribution. In this paper, we use Kolmogorov-Smirnov test to verify that all the datasets $ln r$ of four cities can be reasonably assumed satisfy normal distributions (see Figure \ref{fig:logchangerate} and Table \ref{tab:kstest}). Note here that $l_{t}$ is the length of the flight between time $t-1$ and time $t$. The probability density function of the flight length with the same change variable is log-normal.
\begin{eqnarray}\label{Lognormal}
f(l_{t}) &=& \frac{1}{\sqrt{2 \pi D t} } \frac{1}{l_{t}} exp \left[- \frac{1}{2 D t} (\ln l_{t}- v t) ^2 \right] \nonumber \\
         &=&  \frac{1}{\sqrt{2 \pi D t }} l_{t} ^ {-1+ \frac{v}{D}} exp \left[- \frac{1}{2 
         D t}(\ln ^2 l_{t} + v ^2 t^2) \right]
\end{eqnarray}

 Given
    $$f(l_{t})= \frac{1}{l_{t} \sqrt{2 \pi D t}}exp[- \frac{(\ln l_{t}- v t) ^2}{2 D t}] $$
    $$= \frac{1}{l_{t} \sqrt{2 \pi D t}}exp[- \frac{(\ln l_{t})^2 - 2 v t \ln l_{t} + v ^2 t^2}{2 D t}]$$
    $$= \frac{1}{l_{t} \sqrt{2 \pi D t}}exp[- \frac{(\ln l_{t})^2 + v ^2 t^2}{2 D t}] exp( \frac{v  \ln l_{t}}{ D })$$
    
    $$= \frac{1}{l_{t} \sqrt{2 \pi D t}}l_{t} ^ { \frac{v}{D}} exp[- \frac{(\ln l_{t})^2 + v ^2 t^2}{2 D t}]$$
    
    $$= \frac{1}{\sqrt{2 \pi D t}}l_{t} ^ {-1+ \frac{v}{D}} exp[- \frac{(\ln l_{t})^2 + v ^2 t^2}{2 D t}]$$
        
    This form shows that the log-normal distribution can be mistaken for an apparent power law. If $\sigma \to \infty$, then $\frac{(\ln l_{t})^2}{2 D t} \to 0$. $$f(l_{t}) \to \frac{1}{ \sqrt{ 2 \pi D t}} exp [- \frac{v ^2 t}{2 D}] l_{t} ^{-1+ \frac{v}{D}} \to C l_{t}^ \alpha$$ 
    
    The Probability Density Function of log-normal distribution is indistinguishable from that of power law distribution $f(l_{t})=Cl_{t}^{- \alpha }$, where $ 1 < \alpha \le 3$. 
    
    If there exists a lower bound $l_{min}$, 
    $$l_{t} = max(l_{min}, r_{t-1}l_{t-1})$$
    then $L_{t}$ converges to a power law distribution, log-normal easily pushed to a power law model.
    
    Here the $v$ and the $D$ are the normalized mean and variance of $\ln R$.

If there exists a lower bound $l_{min}$, such that $l_{t} = max(l_{min}, r_{t-1}l_{t-1})$, 
then the random variable $L_{t}$ converges to a power law distribution, log-normal easily pushed to a power law model.

\section*{Discussion}
Previous research suggests that power laws widely exist in city population, financial markets and city-size \cite{luckstead2017pareto,bee2019distribution}. However, the rank-size distribution between cities \cite{wu2019transit} is mostly static, The dynamic urban power-law distribution focuses on the change of specific indicators over time, while the systematic change \cite{chroneer2019urban} among urban factors has not been studied. By using a recently popular neural network embedding technique to reduce the dimension of urban factor data-sets into two dimensions: economy and society, we explore the city development trajectory of Hong kong, Shanghai, Singapore and Tokyo. The urban development of Hong Kong tends to be truncated power law distribution. This is probably because the rapid development of China's reform and opening up has weakened Hong Kong's status as an important city in Southeast Asia, and Hong Kong is no longer the uniquely preferred city in the allocation of various resources in China.

\section*{Methods}  
\textbf{Data Sets.} 
We collected the official data-sets of Hong kong (see Table \ref{tab:HKdatasets}), Shanghai (see Table \ref{tab:SHAdatasets}), Singapore (see Table \ref{tab:SGdatasets}) and Tokyo (see Table \ref{tab:TYOdatasets}) in our work, The data-sets of the four cities were collated and matched. Using the embedding technique to reduce the dimensions of those data-sets into two dimensions: economy and society. 
Then, we draw the urban development trajectory (see Figure \ref{fig:Anto-Encoder}) with economy as $x$-coordinates and society as $y$-coordinate. 
we extract the following information from the graph: flight lengths.

\textbf{Obtaining Flight Length of each factor.} To the best of our knowledge, this article is the first work that examines the flight length distribution of urban development. Firstly, we get raw data of each year's flight length for each factor. The GDP factor ranges from dozens to thousands, while Proportion of industry ranges between 0 and 4, the range of values of raw data varies widely. To avoid the flight length being governed by large value data, we use min-max normalization to scale the range  of each factor in [0, 1].

\textbf{Obtaining Flight length of urban development by Embedding.} The Manifold Hypothesis states that real-world high-dimensional data lie on low-dimensional manifolds embedded within the high-dimensional space.  In this paper, we try to get embedding layer through training Auto-encoder (AE), which is a type of artificial neural network. Firstly, we classifies the data into two classes, for example, regarding GDP, per capita GDP and primary GDP as economic variables, regarding population and General higher education as social variables. Secondly, we use min-max scaling to make sure variables that are measured at different scales contribute equally to the model fitting. Thirdly, in the AE, the input feature,the dataset of economic variables or social variables, is transformed into one latent space with the encoder and then reconstructed from latent space with decoder. The encoder is used as a dimensionality reducer. To train this AE, an Adam algorithm was applied as an optimizer and mean square error (MSE) as a loss function. 
We use two-layer fully connected networks as the encoder and decoder, and the loss function is
\begin{eqnarray}
    \mathcal{L}(\textbf x, \textbf x') &=& \left \| \textbf x - \textbf x' \right \|  ^2 \nonumber \\
    &=& \left \| \textbf x - f(h(\textbf x)) \right \| ^2
\end{eqnarray}
where $\textbf x \in \mathbb{R} ^ n $ is the input feature of one year and $n$ is the number of variables. The data from $m$ years construct $m$ training samples. $\textbf x'$ is the output of the decoder. The computation of the encoder and decoder is defined as 
\begin{eqnarray}
    h(\textbf {x}) &=& \sigma ( W_2 \sigma ( W_1 \textbf x + b_1 ) + b_2 ), \nonumber \\
    f( \textbf y) &=& \sigma ' (W_2' \sigma' (W_1' \textbf y + b_1') +b_2')
\end{eqnarray}
where $ W_1, W_2, W_1', W_2', b_1, b_2, b_1', b_2'$ are learnable parameters of the network, and $\sigma, \sigma'$ are the activation functions. We train the AE by minimizing the MSE of the input feature and the output of the decoder. And the output of the encoder $h( \mathbf x )$ is the embedding of the original input. 

\textbf{Identifying the Scale Range.} 
To fit a heavy tailed distribution such as a power law distribution, 
we need to determine what portion of the data to fit $x_{min}$ and 
the scaling parameter $\alpha$. We use the methods from \cite{clauset2009power} to determine $x_{min}$ and $\alpha$. We create a power law fit starting from each value in the dataset. Then we select the one that results in the minimal Kolmogorov-Smirnov distance between the data and the fit, as the optimal value of $x_{min}$. After that, the scaling parameter $\alpha$ in the power law distribution is given by

\begin{eqnarray}
    \alpha &=& 1 + n \left( \sum_{i=1}^{n} \ln \frac{x_{i}}{x_{min}} \right) ^{-1}
\end{eqnarray}
where $x_{i}$ are the observed values of $x_{i}>x_{min}$ and $n$ is the number of samples.

\textbf{Exponential transformation} The probability density function of exponential distribution can be transformed into power law distribution. Let $X$ be an exponential random variable whose probability density function is given by $ P(X=x) = \lambda e^{- \lambda x}, \lambda>0, x>0$, then the cumulative probability function is given by
\begin{eqnarray}
    P(X \le x) &=& \int_{0}^x \lambda e^{- \lambda t} dt \nonumber \\
    &=& 1-e^{- \lambda x},     \lambda>0, x>0
\end{eqnarray}
and let $Y$ be the random variable obtained through the transformation $Y=ke^{X}$, $k>0$, we can express the cumulative density function of $Y$  in terms of the cumulative density function of $X$ as
\begin{eqnarray}
P(Y \le y) &=& P(ke^{X} \le y)  \nonumber \\
&=& P \left[ X \le \ln (\frac{y}{k}) \right]  \nonumber \\
&=& 1 - e^{- \lambda \ln (\frac{y}{k})}  \nonumber \\
&=& 1 - (\frac{y}{k})^{- \lambda}   \nonumber   \\
&=& 1 - k^ \lambda y^ {- \lambda}   \nonumber \\
P(Y = y)  &=& \lambda k^{ \lambda } y ^{-( 1+ \lambda)}  
\end{eqnarray}
which corresponds to the Probability Density Function of the Power-law distribution with shape factor $\alpha = 1 + \lambda $.

\textbf{Akaike weights.} 
We use Akaike weights to choose the best fitted distribution. An Akaike weight is a normalized distribution selection criterion.
Its value is between 0 and 1. A larger value indicates a better fitted distribution.  

Akaike's information criterion (AIC) is used in combination with Maximum likelihood estimation (MLE). 
MLE finds an estimator of $\hat{\theta}$ that maximizes the likelihood function $L(\hat{\theta}|data)$ 
of one distribution. 
AIC is used to describe the best fitting one among all fitted distributions, 
\begin{eqnarray} 
    AIC &=& -2 log \left(L(\hat{\theta}|data)\right) + 2K.
\end{eqnarray}
Here $K$ is the number of estimable parameters in the approximating model.

After determining the AIC value of each fitted distribution, we normalize these values as follows. 
First of all, we extract the difference between different AIC values called $\Delta_i$, 
\begin{eqnarray}
    \Delta_i &=& AIC_i - AIC_{min}. 
\end{eqnarray}

Then Akaike weights $W_i$ are calculated as follows,
\begin{eqnarray} 
    W_i &=& \frac{exp(-\Delta_i / 2)}{\sum_{r = 1}^{R} exp(-\Delta_i / 2)}.
\end{eqnarray}
The statistics can be see in Table \ref{tab:Akaikeweights}.


A List of abbreviations

HK: Hong Kong

SHA: Shanghai

SG: Singapore

TYO: Tokyo

pdf: probability density function

SMP: Stochastic Multiplicative Processes

AE: Auto-encoder

MSE: mean square error

AIC: Akaike's information criterion

MLE: maximum likelihood estimation


\begin{backmatter}

\section*{Availability of data and  material}
    The data can be collected from official websites, which are listed in Table \ref{tab:HKdatasets}, Table \ref{tab:SHAdatasets}, Table \ref{tab:SGdatasets} and Table \ref{tab:TYOdatasets}.
    
\section*{Funding}  
This work is partially supported by
National Natural Science Foundation of China (Grant No. 61972286).

\section*{Competing interests}
  The authors declare that they have no competing interests.

\section*{Author's contributions}
    Weixiong Rao and Kai Zhao conceived the experiments,  Linfang Tian and Jiamin Yin conducted the experiments, Linfang Tian analysed the results.  All authors reviewed the manuscript. 
    
\section*{Acknowledgements}
Not applicable.


\bibliographystyle{bmc-mathphys} 
\bibliography{bmc}      
      
\newpage
\section*{Figures and Tables}
    \begin{figure}[h]
    \subfigure[HK Auto-Encoder]{
        \includegraphics[scale=0.28]{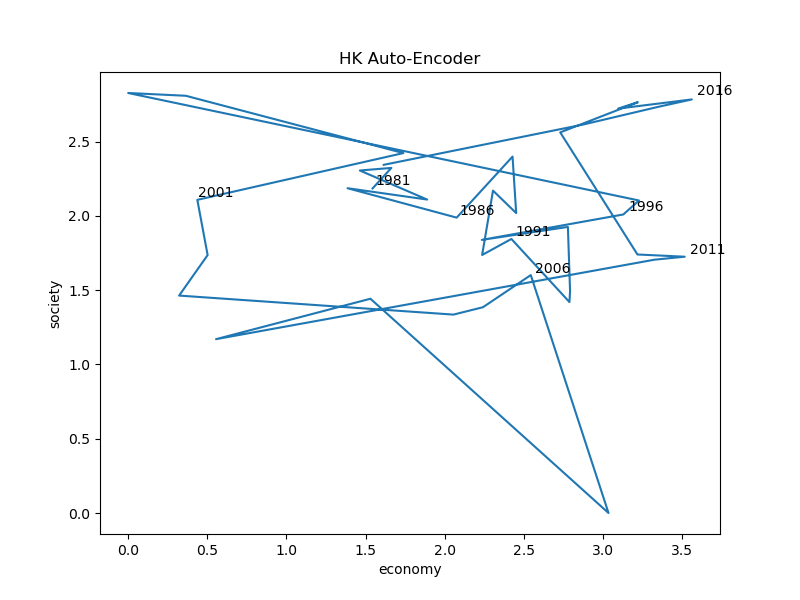}
        }
    \subfigure[SHA Auto-Encoder]{
        \includegraphics[scale=0.28]{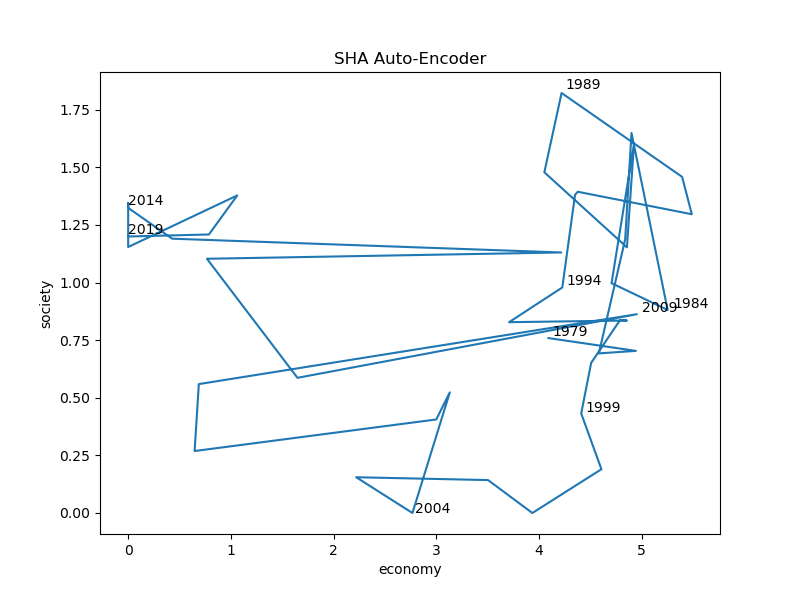}
        }
    \subfigure[SG Auto-encoder]{
       \includegraphics[scale=0.28]{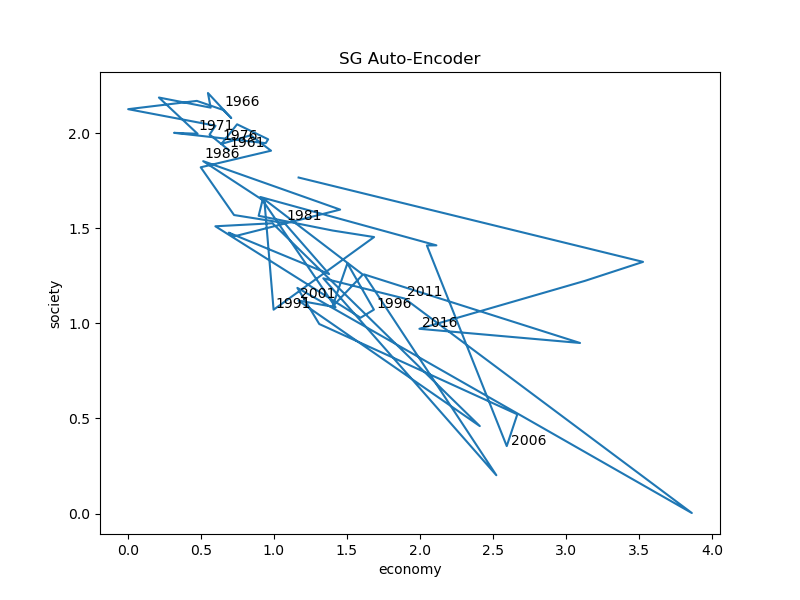}
       }
    \subfigure[TYO Auto-Encoder]{
       \includegraphics[scale=0.28]{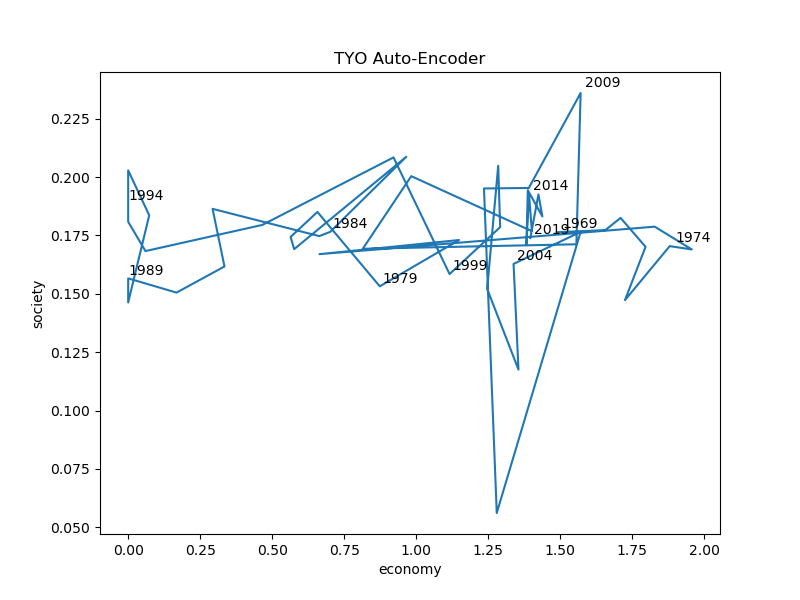}
       }
    \caption{\csentence{Auto-Encoder}    The economical and social trajectory of urban development.As can be seen from the figure(a), before the opening up of the Pudong District in 1990, the development step size of Shanghai was relatively small.The 2010 Shanghai World Expo has led to a rapid economic development. As can be seen from the figure(b), after Hong Kong's return to China in 1997, it enjoyed social and economical stability and prosperity, and successfully fended off the Asian financial crisis in 1998. After 2010, Hong Kong's economic and social development was sluggish.}
    \label{fig:Anto-Encoder}
    \end{figure}

    \begin{figure}[h]
    \subfigure[HK histogram]{
    \includegraphics[scale=0.33]{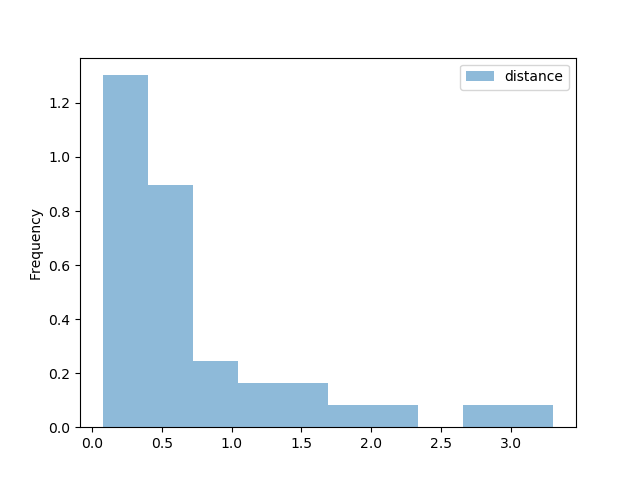}
    }
    \subfigure[SHA histogram]{
    \includegraphics[scale=0.33]{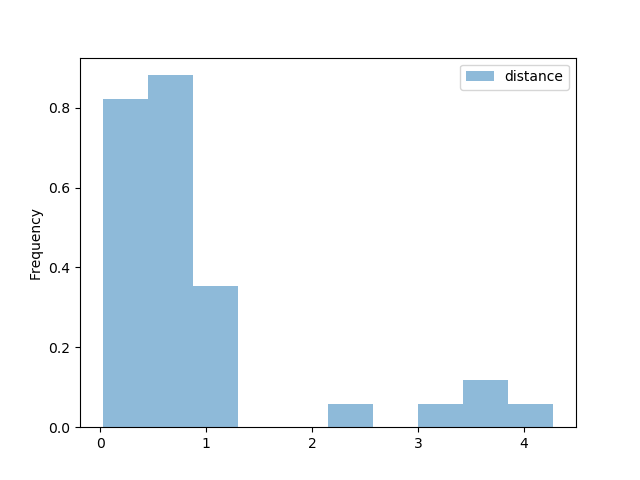}
    }
    \subfigure[SG histogram]{
    \includegraphics[scale=0.3]{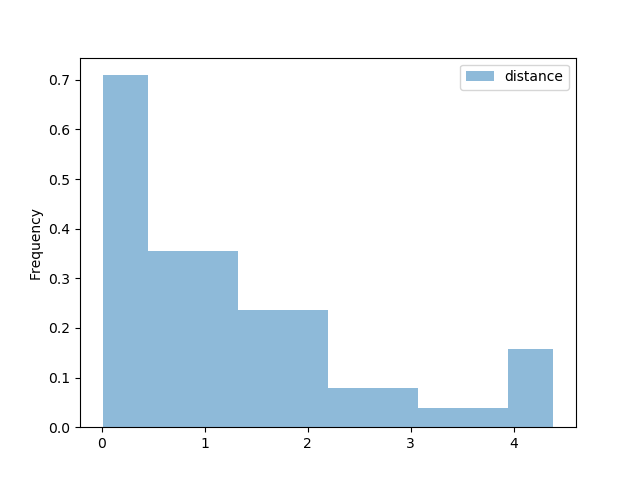}
    }
    \subfigure[TYO histogram]{
    \includegraphics[scale=0.33]{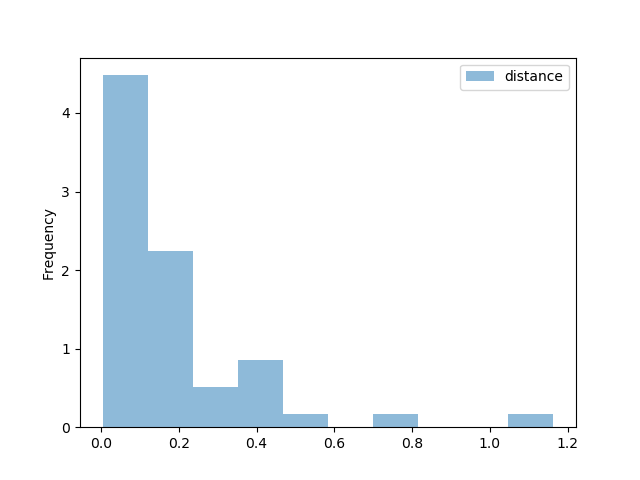}
    }
    \caption{\csentence{Histogram of flight length}.As can be seen from both the figure(a), (b), (c) and (d), all of the histograms are positive skewed. The statistical analysis of the step size of the random walk shows obvious "heavy tail" feature, which satisfies the walking characteristics of frequent short-distance walk and occasional long-distance jump.}
    \label{fig:histogram}
    \end{figure}

    \begin{figure}[h]
    \subfigure[HK Truncated power-law]{
    \includegraphics[scale=0.39]{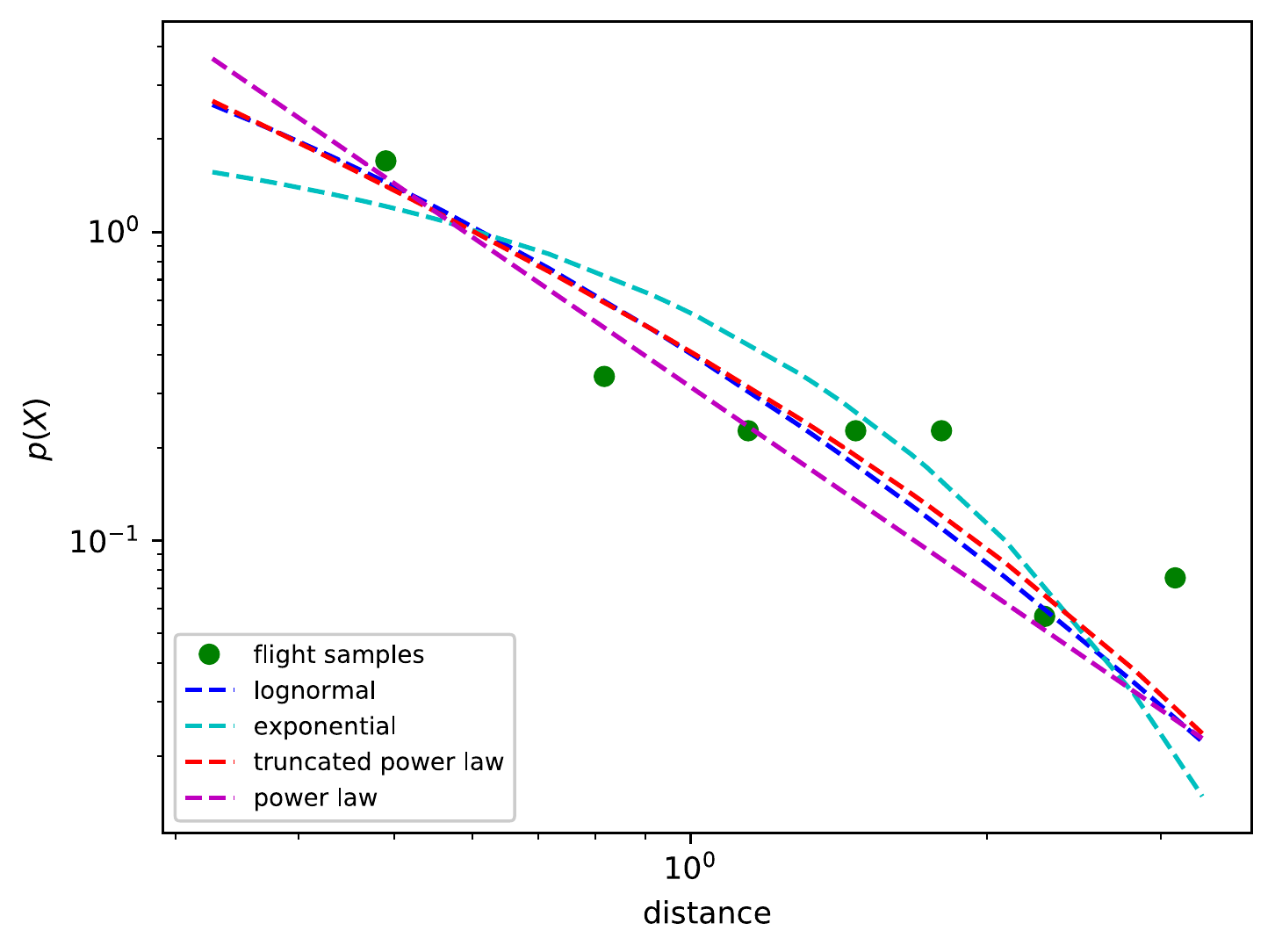}
    }
    \subfigure[SHA Power-law]{
    \includegraphics[scale=0.39]{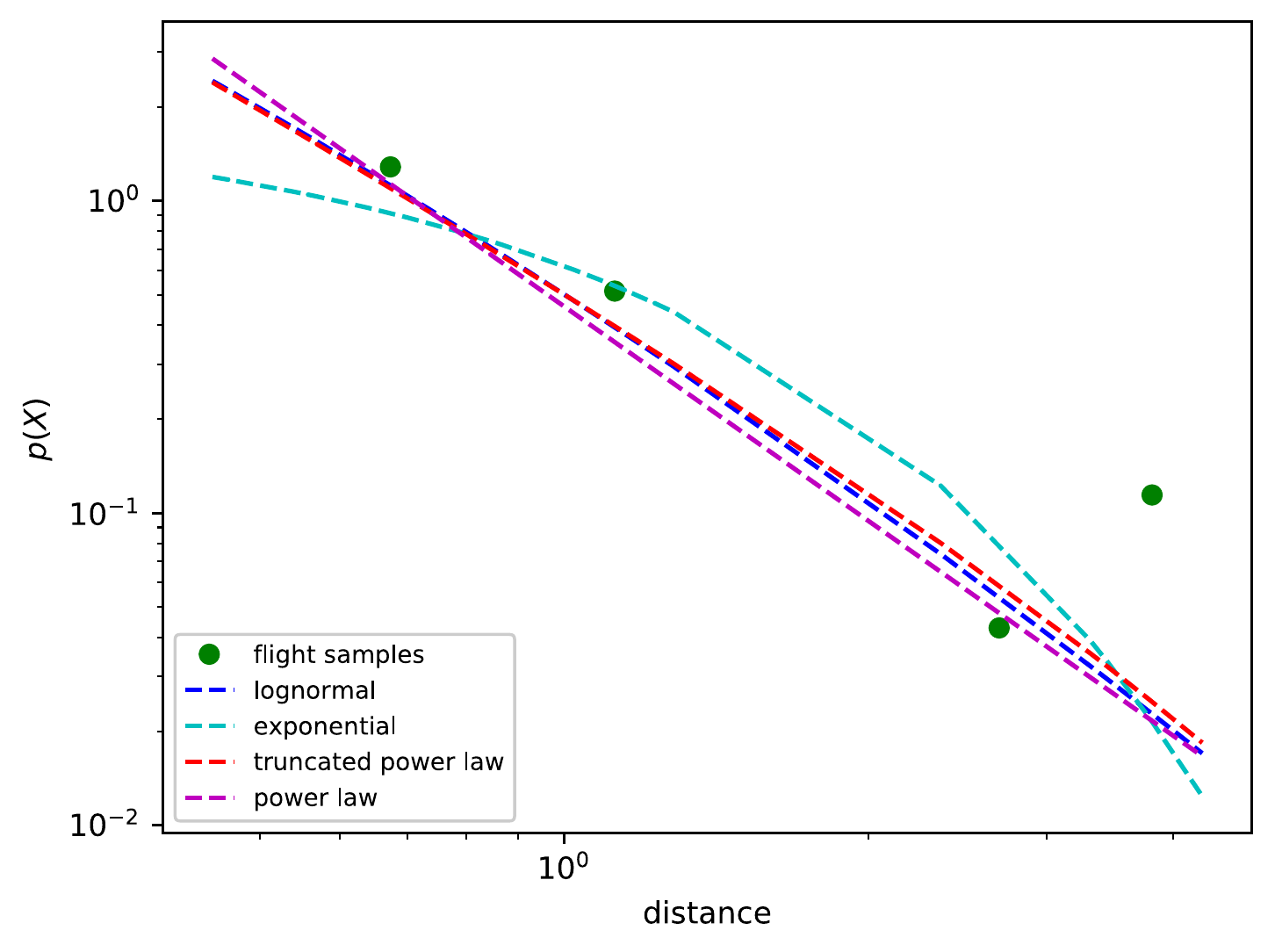}
    }
    \subfigure[SG Exponential]{
    \includegraphics[scale=0.39]{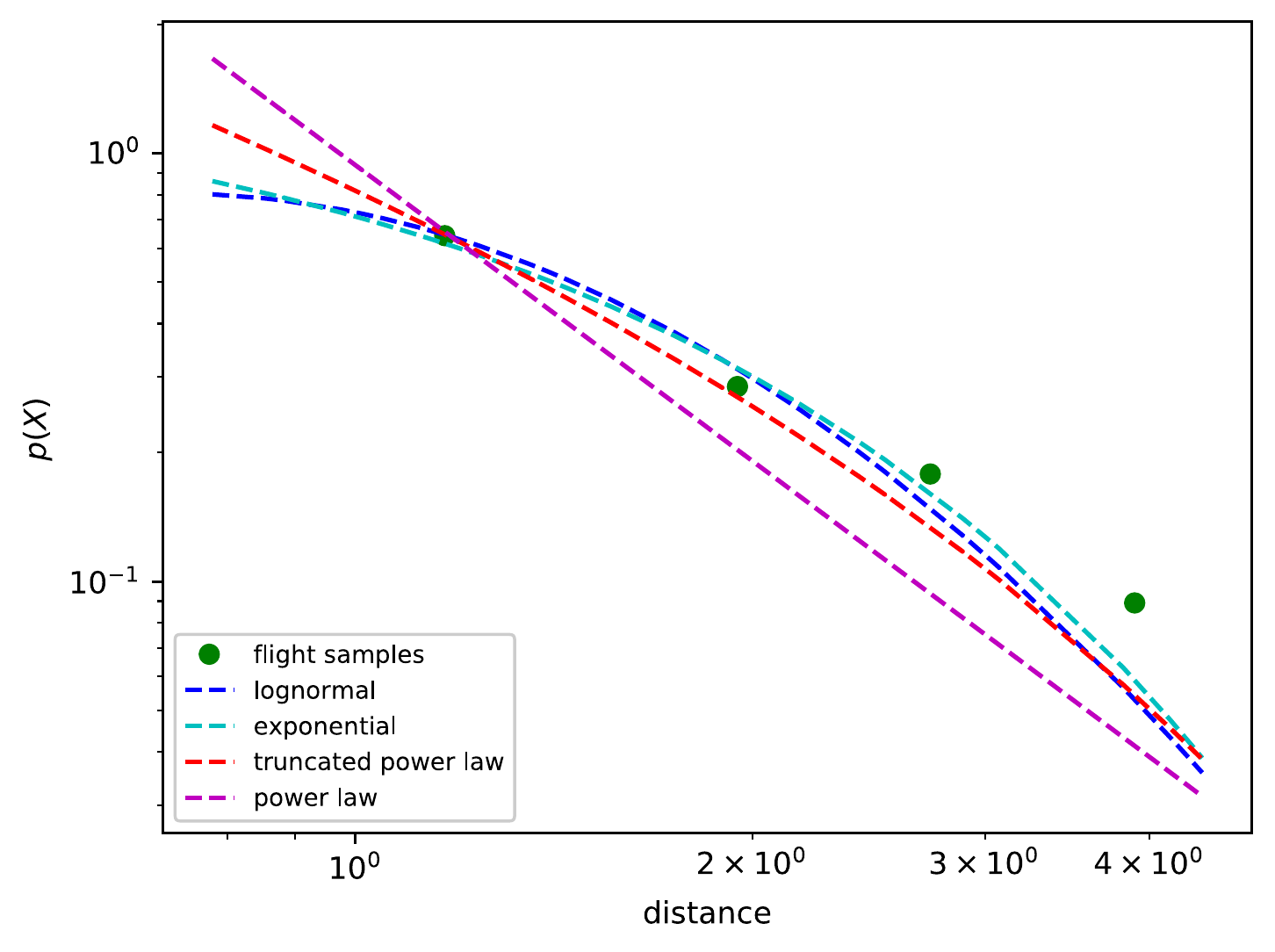}
    }
    \subfigure[TYO Power-law]{
    \includegraphics[scale=0.39]{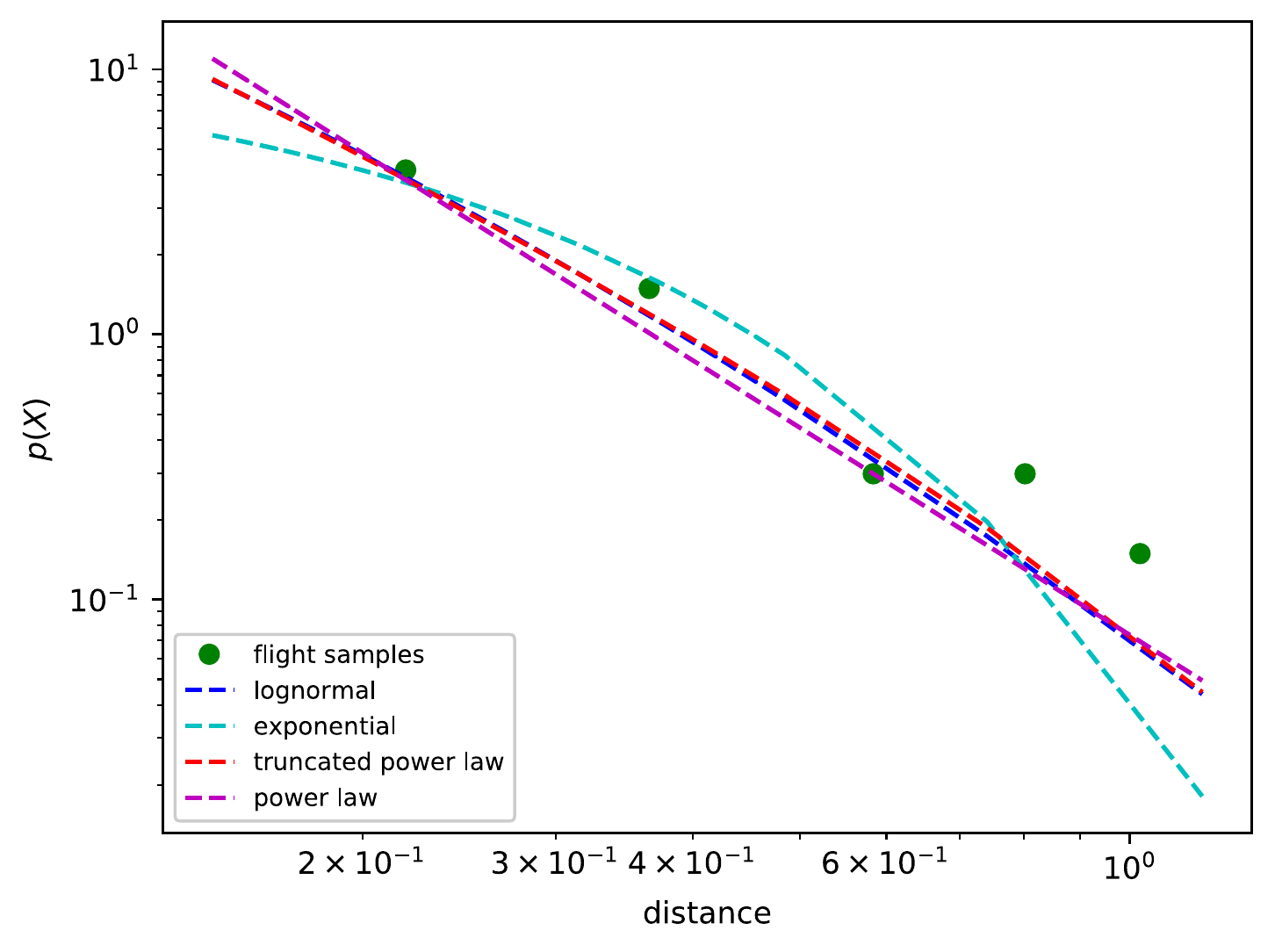}
    }
    \caption{\csentence{Fitted distributions}. Hong Kong has experienced rapid development and sharp decline. The urban development step size is relatively rich, with more long-distance jump. The fitted Truncated power-law distribution has a thicker tail, with alpha 1.3547. The development step size of Shanghai has more short-distance walk and less long-distance jump. It is still in a stable development period. The $\alpha$ of the fitted Power-law distribution is 2.2829. Singapore has been developing with relatively constant multiplicative factor, the fitted Exponential distribution with $\lambda = 1.6075$. With the Exponential Transformation, the exponential distribution can deducted to Power-law distribution with $\alpha = 1+ \lambda = 2.6075$. Tokyo has been developing relatively earlier than Shanghai, the fitted Power-law distribution has a larger $\alpha$ 2.6016.}
    \label{fig:Power-law}
    \end{figure}

    \begin{figure}[h]
    \subfigure[HK r value]{
    \includegraphics[scale=0.36]{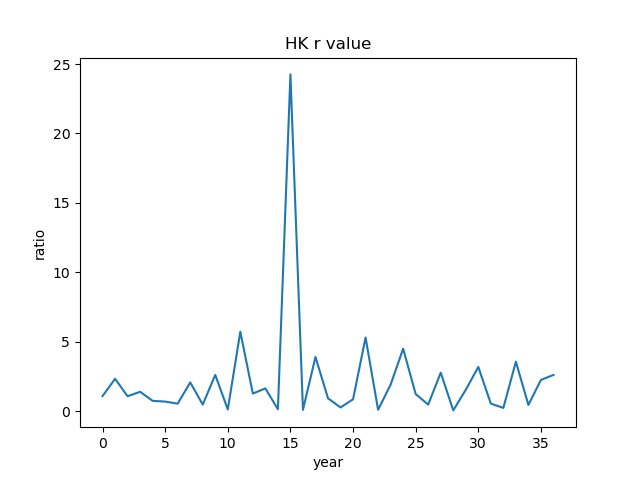}
    }
    \subfigure[SHA r value]{
    \includegraphics[scale=0.36]{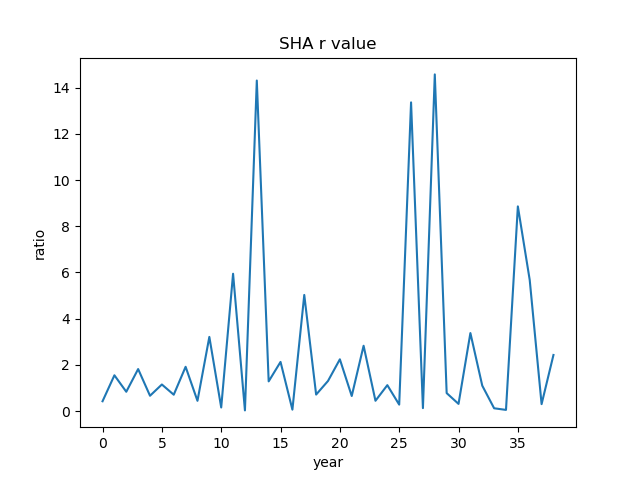}
    }
    \subfigure[SG r value]{
    \includegraphics[scale=0.36]{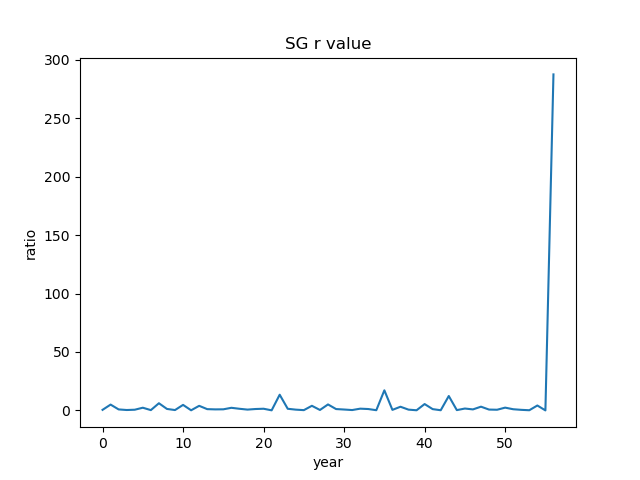}
    }
    \subfigure[TYO r value]{
    \includegraphics[scale=0.36]{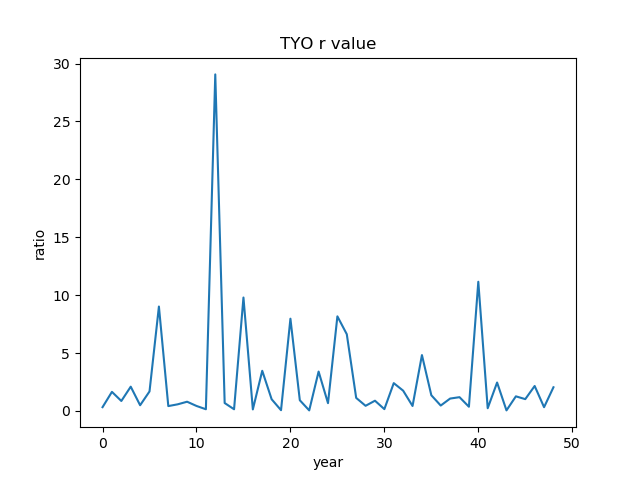}
    }
    \caption{\csentence{The relative change rates}. The change rate is defined as the relative change of length between two consecutive flights. From these figures we observe that the change rate are uncorrelated from one time interval to the other.}
    \label{fig:change rate}
    \end{figure}

    \begin{figure}[h]
    \subfigure[HK $\ln r$ value]{
    \includegraphics[scale=0.36]{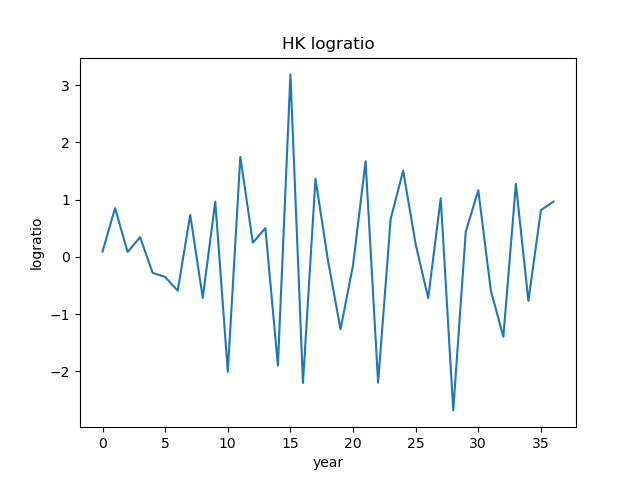}
    }
    \subfigure[SHA $\ln r$ value]{
    \includegraphics[scale=0.36]{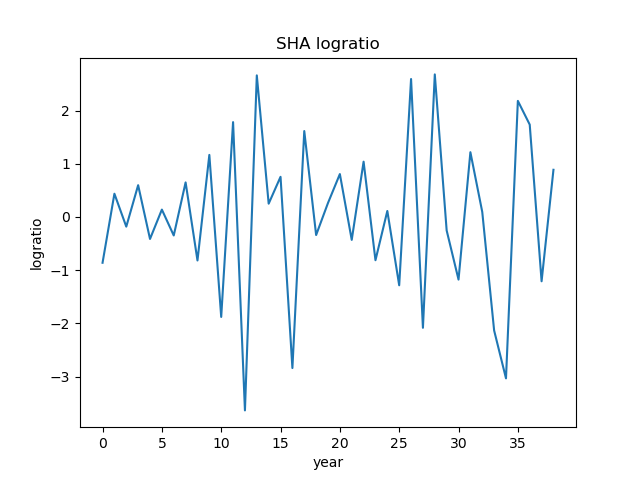}
    }
    \subfigure[SG $\ln r$ value]{
    \includegraphics[scale=0.36]{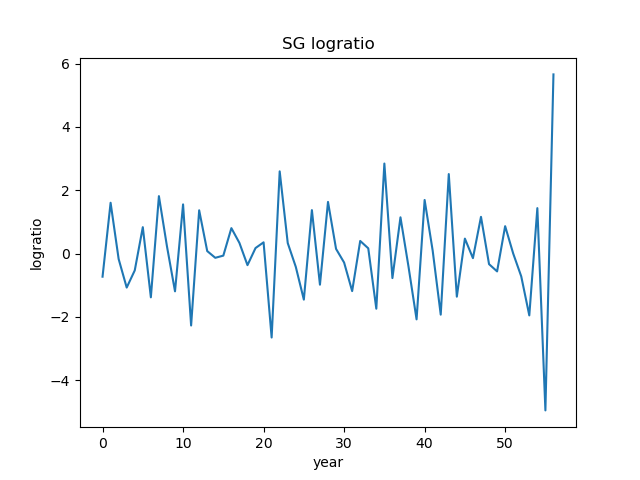}
    }
    \subfigure[TYO $\ln r$ value]{
    \includegraphics[scale=0.36]{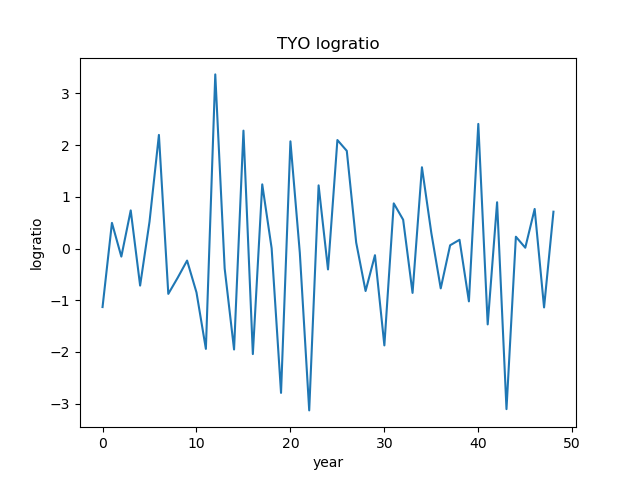}
    }
    \caption{\csentence{$\ln r$ of four cities}}
    \label{fig:logchangerate}
    \end{figure}

\begin{table}[h]
\caption{The analysis of Hongkong, Shanghai, Singapore and Tokyo Datasets}
    \begin{tabular}{c c| c c c c}
    \hline
        \textbf{Class}      & \textbf{Description}   &\textbf{HK}    &\textbf{SHA}   &\textbf{SG} &\textbf{TYO} \\  \hline\hline
        \textbf{ Economy}    &GDP   &1981-2019   &1978-2018   &1960-2019  &1968-2019  \\  \hline
            &Primary industry   &1981-2019   &1978-2018   &NA  &NA   \\   \hline
            &Secondary industry   &1981-2019   &1978-2018 &1960-2019  &NA   \\  \hline
            &Tertiary industry   &1981-2019   &1978-2018  &1960-2019  &NA   \\  \hline
            &Share of Primary industry  &1981-2019   &1978-2018 &NA  &NA \\  \hline
            &Share of Secondary industry   &1981-2019   &1978-2018 &1960-2019  &NA  \\  \hline
            &Share of Tertiary industry &1981-2019   &1978-2018 &1960-2019  &NA  \\  \hline
            &Per capita GDP   &1981-2019   &1978-2018  &NA  &NA  \\  \hline
            &Government revenue   &1981-2019   &1978-2018 &NA &NA  \\  \hline
            &Government expenditure   &1981-2019   &1978-2018  &1960-2019 &NA  \\  \hline
            &Personal income  &NA   &NA  &NA  &1960-2019   \\ \hline
            &Original insurance income   &NA   &1978-2018 &NA  &NA \\  \hline
            &Original insurance pays out   &NA   &1978-2018  &NA &NA  \\  \hline
            &Total fixed asset investment   &NA   &1978-2018 &1960-2019 &NA  \\  \hline
            &Industry   &NA   &1978-2018  &1960-2019   &1968-2019  \\  \hline
            &GDP per capita(Dollar)   &NA   &1978-2018  &NA  &NA \\  \hline
            &Proportion of industry   &NA   &1978-2018  &1960-2019 &NA \\  \hline
            &Gross agricultural production   &NA   &1978-2018  &NA &NA \\  \hline
            &Gross industrial production   &NA   &1978-2018 &1960-2019  &1968-2019  \\  \hline
        \textbf{Society}            &Population  &1981-2019  &1978-2018 &1960-2019  &1968-2019  \\\hline
            &Labor    &NA   &NA   &NA     &1968-2019 \\  \hline
            &General Tertiary education  &1981-2019  &1978-2018  &1960-2019   &1968-2019 \\ \hline
            &Ordinary secondary school  &1981-2019  &1978-2018  &1960-2019   &1968-2019 \\ \hline
            &Ordinary primary school  &1981-2019  &1978-2018  &1960-2019 &1968-2019 \\ \hline
            &Book print run  &NA  &1978-2018  &NA  &NA  \\ \hline
            &Journal print run  &NA  &1978-2018  &NA &NA \\ \hline
            &Newspaper print run  &NA  &1978-2018  &NA &NA  \\ \hline
    \end{tabular}
    \label{tab:4citiesdatasets}
\end{table}

\begin{table}[th]
\caption{Fitted distributions. With $ 1 < \alpha \le 3$, the Power-law distribution has infinite variance.  It has infinite mean as $ 1 < \alpha \le 2$ and finite mean as $2 < \alpha \le 3 $.}
    \begin{tabular}{ c c }
    \hline
   \textbf{ Distribution} &  \textbf{Probability density function (pdf)} \\ \hline\hline
    Exponential & $\lambda  e ^{- \lambda x}$ \\
    Power-law & $C x ^{- \alpha }$ \\ 
    Lognormal & $ \frac{1}{x \sigma \sqrt{2 \pi }} exp[- \frac{( \ln (x) -  \mu )^2}{2 \sigma ^2} ]$ \\
    Truncated power-law & $C x ^{- \alpha} e^{- \gamma x}$ \\  \hline
   \end{tabular}
\label{tab:pdf}
\end{table}

\begin{table}[th]
\caption{Akaike weights of fitted distributions in the four cities datasets.}
    \begin{tabular}{c c c c c}
    \hline
    \textbf{Cities} & \textbf{Exponential} & \textbf{Power-law} & \textbf{Lognormal} & \textbf{Truncated Power-law} \\ \hline\hline
    HK & 0.1979 & 0.2568 & 0.2226 & 0.3227 \\
    SHA & 0.0401 & 0.4603 & 0.2163 & 0.2814 \\
    SG & 0.6717 & 0.0014 & 0.1283 & 0.1985 \\
    TYO & 0.1594 &  0.4132 & 0.1979 & 0.2295\\ \hline 
    \end{tabular}
    \label{tab:Akaikeweights}
\end{table}

\begin{table}[h!]
\caption{The calculated and estimated parameters for consecutive flights length in the four cities datasets, with $\ln R_{i}$ taken in the interval [0.48, 1.48]\cite{sornette1997convergent}. The mean is noted as $v'$, and variance is $D'$, $\hat{\alpha}$ is calculated, and $\alpha$ is the fitted exponent. Here the walk lengths of Hong kong is fitted Truncated power-law rather than Power-law distribution.}
    \begin{tabular}{c c c c c c c}
    \hline
     \textbf{Cities} & $l_{min}$ & $v'$ & $D'$ & $v'/D'$ & $\hat{\alpha}$ & $\alpha$ \\ \hline\hline
     HK* & 0.0752 & -0.1904 & 0.1443 & -1.3200 & 2.3200 & 1.3547     \\
     SHA & 0.0216 & -0.1010 & 0.0751 & -1.3453 & 2.3453 & 2.2829   \\
     SG  & 0.0262 & -0.1310 & 0.1145 & -1.1445 & 2.1445 & 2.6075    \\
     TYO & 0.0033 & -0.11716 & 0.0875 & -1.3390 & 2.3390 & 2.6016    \\ \hline
    \end{tabular}
    \label{tab:estimatedparameters}
\end{table}

\begin{table}[th]
\caption{The $p$ value of Kolmogorov-Smirnov test for four city datasets.}
    \begin{tabular}{c c}
    \hline
     \textbf{Cities} & $p$ \textbf{value} \\ \hline\hline
     HK & 0.9804 \\
     SHA & 0.9477 \\
     SG  & 0.7399 \\
     TYO & 0.9933 \\ \hline
    \end{tabular}
    \label{tab:kstest}
\end{table}

\begin{table}[th]
\caption{The URL of HK dataset.}
    \begin{tabular}{ c c }
    \hline
    \textbf{Description}  & \textbf{URL} \\ \hline \hline
    GDP  & \url{https://www.censtatd.gov.hk/sc/web_table.html?id=31#}   \\ \hline
    Per capita GDP & \url{https://www.censtatd.gov.hk/sc/web_table.html?id=31#}   \\ \hline
    Primary industry & \url{https://www.censtatd.gov.hk/sc/web_table.html?id=35#}   \\ \hline
    Secondary industry & \url{https://www.censtatd.gov.hk/sc/web_table.html?id=35#}   \\ \hline
    Tertiary industry	& \url{https://www.censtatd.gov.hk/sc/web_table.html?id=35#}   \\ \hline
    Proportion of primary industry & \url{https://www.censtatd.gov.hk/sc/web_table.html?id=36#}   \\ \hline
    Proportion of secondary industry & \url{https://www.censtatd.gov.hk/sc/web_table.html?id=36#}   \\ \hline
    Proportion of tertiary industry & \url{https://www.censtatd.gov.hk/sc/web_table.html?id=36#}   \\ \hline
    Government revenue & \url{https://www.censtatd.gov.hk/sc/web_table.html?id=193#}   \\ \hline
    Government expenditure & \url{https://www.censtatd.gov.hk/sc/web_table.html?id=194#}   \\ \hline
    Population & \url{https://www.censtatd.gov.hk/sc/web_table.html?id=1A#}   \\ \hline
    Labour force & \url{https://www.censtatd.gov.hk/sc/web_table.html?id=6#}   \\ \hline
    Primary school & \url{https://www.censtatd.gov.hk/sc/scode370.html#section6}   \\ \hline 
    Secondary school & \url{https://www.censtatd.gov.hk/sc/scode370.html#section6}    \\ \hline
    University & \url{https://www.censtatd.gov.hk/sc/scode370.html#section6}   \\ \hline 
    \end{tabular}
\label{tab:HKdatasets}
\end{table}

\begin{table}[th]
\caption{The URL of SHA dataset.}
    \begin{tabular}{ c c }
    \hline
    \textbf{Description}  & \textbf{URL} \\ \hline \hline
    GDP  & \url{http://tjj.sh.gov.cn/tjnj/nj20.htm?d1=2020tjnj/C0401.htm} \\ \hline
    Primary industry  & \url{http://tjj.sh.gov.cn/tjnj/nj20.htm?d1=2020tjnj/C0401.htm} \\ \hline
    Secondary industry  & \url{http://tjj.sh.gov.cn/tjnj/nj20.htm?d1=2020tjnj/C0401.htm} \\ \hline
    Tertiary industry  & \url{http://tjj.sh.gov.cn/tjnj/nj20.htm?d1=2020tjnj/C0401.htm} \\ \hline
    Industry  & \url{http://tjj.sh.gov.cn/tjnj/nj20.htm?d1=2020tjnj/C0401.htm} \\ \hline
    General public budget revenue & \url{http://tjj.sh.gov.cn/tjnj/nj20.htm?d1=2020tjnj/C0501.htm}  \\ \hline
    General public budget expenditure &  \url{http://tjj.sh.gov.cn/tjnj/nj20.htm?d1=2020tjnj/C0501.htm}  \\ \hline
    Proportion of primary industry &  \url{http://tjj.sh.gov.cn/tjnj/nj20.htm?d1=2020tjnj/C0404.htm}  \\ \hline
    Proportion of Secondary industry &  \url{http://tjj.sh.gov.cn/tjnj/nj20.htm?d1=2020tjnj/C0404.htm}  \\ \hline
    Proportion of Tertiary industry & \url{http://tjj.sh.gov.cn/tjnj/nj20.htm?d1=2020tjnj/C0404.htm}  \\ \hline
    Proportion of industry & \url{http://tjj.sh.gov.cn/tjnj/nj20.htm?d1=2020tjnj/C0404.htm}  \\ \hline
    Total fixed asset investment & \url{http://tjj.sh.gov.cn/tjnj/nj20.htm?d1=2020tjnj/C0701.htm}   \\ \hline
    General public budget revenue	& \url{http://tjj.sh.gov.cn/tjnj/nj20.htm?d1=2020tjnj/C0501.htm}   \\ \hline
    General public budget expenditure & \url{http://tjj.sh.gov.cn/tjnj/nj20.htm?d1=2020tjnj/C0501.htm}   \\ \hline
    Gross agricultural production & \url{http://tjj.sh.gov.cn/tjnj/nj20.htm?d1=2020tjnj/C1201.htm}   \\ \hline
    Gross industrial production &
        \url{http://tjj.sh.gov.cn/tjnj/nj20.htm?d1=2020tjnj/C1301.htm}   \\ \hline
    Original insurance income	& \url{http://tjj.sh.gov.cn/tjnj/nj20.htm?d1=2020tjnj/C1801.htm}   \\ \hline
    Original insurance pays out & \url{http://tjj.sh.gov.cn/tjnj/nj20.htm?d1=2020tjnj/C1801.htm}   \\ \hline
    Resident population at year-end & \url{http://tjj.sh.gov.cn/tjnj/nj20.htm?d1=2020tjnj/C0201.htm}   \\ \hline
    Registered population at year-end & \url{http://tjj.sh.gov.cn/tjnj/nj20.htm?d1=2020tjnj/C0201.htm}   \\ \hline
    General higher education & \url{http://tjj.sh.gov.cn/tjnj/nj20.htm?d1=2020tjnj/C2103.htm}   \\ \hline
    Ordinary secondary school & \url{http://tjj.sh.gov.cn/tjnj/nj20.htm?d1=2020tjnj/C2103.htm}   \\ \hline
    Ordinary primary school & \url{http://tjj.sh.gov.cn/tjnj/nj20.htm?d1=2020tjnj/C2103.htm}   \\ \hline
    Book print run & \url{http://tjj.sh.gov.cn/tjnj/nj20.htm?d1=2020tjnj/C2316.htm}   \\ \hline
    Journal print run & \url{http://tjj.sh.gov.cn/tjnj/nj20.htm?d1=2020tjnj/C2317.htm}   \\ \hline
    Newspaper print run & \url{http://tjj.sh.gov.cn/tjnj/nj20.htm?d1=2020tjnj/C2318.htm}   \\ \hline
   \end{tabular}
\label{tab:SHAdatasets}
\end{table}

\begin{table}[th]
\caption{The URL of SG dataset.}
    \begin{tabular}{ c c }
    \hline
    \textbf{Description}  & \textbf{URL} \\ \hline \hline
    GDP & \url{https://tablebuilder.singstat.gov.sg/table/TS/M015241}     \\ \hline
    Goods Producing Industries & \url{https://tablebuilder.singstat.gov.sg/table/TS/M015241}    \\ \hline
    Services Producing Industries & \url{https://tablebuilder.singstat.gov.sg/table/TS/M015241}    \\ \hline
    Goods Proportioin & \url{https://tablebuilder.singstat.gov.sg/table/TS/M015241}    \\ \hline
    Services Proportion & \url{https://tablebuilder.singstat.gov.sg/table/TS/M015241}    \\ \hline
    Government Consumption & \url{https://tablebuilder.singstat.gov.sg/table/TS/M015241}    \\ \hline
    Gross Fixed Capital Formation & \url{https://tablebuilder.singstat.gov.sg/table/TS/M015051}    \\ \hline
    Total Population & \url{https://tablebuilder.singstat.gov.sg/table/TS/M810001#!}    \\ \hline
    Government Expenditure On Edu & \url{https://tablebuilder.singstat.gov.sg/table/TS/M850011}    \\ \hline
    Primary Schools & \url{https://tablebuilder.singstat.gov.sg/table/TS/M850011}    \\ \hline
    Secondary Schools & \url{https://tablebuilder.singstat.gov.sg/table/TS/M850011}    \\ \hline
    Tertiary & \url{https://tablebuilder.singstat.gov.sg/table/TS/M850011}    \\ \hline
    Literacy Rate & \url{https://tablebuilder.singstat.gov.sg/table/TS/M850001}    \\ \hline
   \end{tabular}
\label{tab:SGdatasets}
\end{table}

\begin{table}[th]
\caption{The URL of TYO dataset.}
    \begin{tabular}{ c c }
    \hline
    \textbf{Description}  & \textbf{URL} \\ \hline \hline
    Loans &          \url{https://www.toukei.metro.tokyo.lg.jp/tnenkan/2019/tn19q3i015.htm}    \\ \hline
    Manufactured goods & \url{https://www.toukei.metro.tokyo.lg.jp/tnenkan/2019/tn19q3i016.htm}    \\ \hline
    GDP & \url{https://www.toukei.metro.tokyo.lg.jp/tnenkan/2019/tn19q3i016.htm}    \\ \hline
    Prefectural income & \url{https://www.toukei.metro.tokyo.lg.jp/tnenkan/2019/tn19q3i016.htm}    \\ \hline
    Population & \url{https://www.toukei.metro.tokyo.lg.jp/tnenkan/2019/tn19q3i002.htm}    \\ \hline
    Labor & \url{https://www.toukei.metro.tokyo.lg.jp/tnenkan/2019/tn19q3i002.htm}    \\ \hline
    Children and students & \url{https://www.toukei.metro.tokyo.lg.jp/tnenkan/2019/tn19q3i017.htm}    \\ \hline
    Elementary schools & \url{https://www.toukei.metro.tokyo.lg.jp/tnenkan/2019/tn19q3i017.htm}    \\ \hline
    Junior secondary & \url{https://www.toukei.metro.tokyo.lg.jp/tnenkan/2019/tn19q3i017.htm}    \\ \hline
    Senior secondary & \url{https://www.toukei.metro.tokyo.lg.jp/tnenkan/2019/tn19q3i017.htm}    \\ \hline
    Universities & \url{https://www.toukei.metro.tokyo.lg.jp/tnenkan/2019/tn19q3i017.htm}    \\ \hline
    \end{tabular}
\label{tab:TYOdatasets}
\end{table}

\end{backmatter}
\end{document}